\documentclass[11pt,preprint]{aastex}

\usepackage{xspace}
\usepackage{epsfig}

\newcommand{\hi}{\ion{H}{1}\xspace}
\newcommand{\hii}{\ion{H}{2}\xspace}

\newcommand{\etal}{et al.\ }
\newcommand{\n}{NGC~} 

\newcommand{\uu}[1]{$^{#1}$} 
\newcommand{\msun}{M$_{\odot}$\xspace}
\def\vi{\hbox{$V\!-\!I$}} 
\newcommand{\e}[1]{$\times10^{#1}$}
\newcommand{\hst}{{\it HST}\xspace}
\newcommand{\hubble}{{\it Hubble Space Telescope}\xspace} 
\newcommand{\kms}{km~s$^{-1}$\xspace}
\newcommand{\lgs}{LGS~3\xspace}
\newcommand{\wfpc}{WFPC2\xspace}
\newcommand{\hstphot}{HSTphot\xspace}
\newcommand{\onec}{\multicolumn{1}{c}}

\begin{document}

\title{The Star Formation History of LGS~3\footnote{Based on
observations with the NASA/ESA {\it Hubble Space Telescope}, 
obtained at the Space Telescope Science
Institute, which is operated by the Association of Universities for
Research in Astronomy, Inc., under NASA contract No.~NAS5-26555}}

\author{Bryan W. Miller}
\affil{Gemini Observatory, Casilla 603, La Serena, Chile}

\author{Andrew E. Dolphin}
\affil{Kitt Peak National Observatory, 950 N. Cherry
Ave., P.O. Box 26732, Tucson, AZ 85726, USA}

\author{Myung Gyoon Lee,  Sang Chul Kim}
\affil{Astronomy Program, SEES, Seoul National University, Seoul
151-742, Korea}

\and
\author{Paul Hodge}
\affil{Astronomy Department, University of Washington, Seattle, WA
98185, USA}

\slugcomment{To appear in The Astrophysical Journal. Gemini Preprint
\#74.}

\begin{abstract}
We have determined the distance and star formation history of the
Local Group dwarf galaxy \lgs from deep \hubble
WFPC2 observations.  \lgs is intriguing because ground-based
observations showed that, while its stellar population is dominated by
old, metal-poor stars, there is a handful of young, blue stars.
Also, the presence of \hi gas makes this a possible ``transition
object'' between dwarf spheroidal and dwarf irregular galaxies.  The
\hst data are deep enough to detect the horizontal branch and
young main sequence for the first time. A new distance of
$D=620\pm20$~kpc has been measured from the positions of the TRGB, the
red clump, and the horizontal branch.  The mean metallicity of the stars
older than 8~Gyr is $[{\rm Fe/H}] = -1.5\pm0.3$. The most recent
generation of stars has $[{\rm Fe/H}] \approx -1$. For the first few
Gyr the global star formation rate was several times higher than the
historical average and has been fairly constant since then.  However,
we do see significant changes in stellar populations and star
formation history with radial position in the galaxy.  Most of the
young stars are found in the central 63~pc (21$''$), where the star
formation rate has been relatively constant, while the outer parts
have had a declining star formation rate.
\end{abstract}


\section{Introduction}

The relationship between dwarf irregular (dI) and dwarf elliptical or
dwarf spheroidal (dE/dSph) galaxies is both uncertain and very
important to a variety of astrophysical problems (see Ferguson \&
Binggeli 1994; Gallagher \& Wyse 1994; Skillman \& Bender 1995; Mateo
1998).  Since they may be among the earliest galactic systems formed,
their study is relevant to research on stellar evolution, primordial
abundances and chemical evolution, and the formation of larger
galaxies. Dwarf irregulars are characterized by irregular appearances,
\hi masses greater than about $10^6$ \msun, and slowly rising rotation
curves. In contrast, dEs have smoother isophotes, very little \hi
(especially in their central regions), and undetectable rotation.  If
Newtonian gravity correctly describes the gravitational force law at
low accelerations, then both types are dominated by dark matter.
Dwarfs are the most numerous type of galaxy, and in hierarchical
galaxy formation scenarios they are the building blocks of more
massive galaxies.

In nearby groups and clusters of galaxies there is significant
morphological segregation of dwarfs similar to that seen between giant
ellipticals and spirals in dense clusters; the dEs are concentrated
near the giant galaxies or the centers of the clusters while the dIs
are spread throughout the volume of the group or cluster (see Ferguson
\& Sandage 1989; Hodge 1994).  It is not known whether there was a
difference in formation process or if all dwarfs formed as dIs and a
fraction of them evolved into dEs.  Dwarf irregulars are both bluer
and of equal or lower surface brightness to the brighter dEs and so
cannot evolve passively into them.  Dwarf irregulars could become
bright dEs if they first go through a blue compact dwarf (BCD) phase
(Bothun \etal 1986).  Recent evidence that BCDs are more compact than
dIs of similar brightnesses (Salzer \& Norton 1999) suggests that BCDs
may be a special type of compact dI; their underlying stellar
populations have higher central surface brightnesses and smaller scale
lengths than regular dIs. Alternatively, the process that produces
that starbursts in BCDs could be effective at transferring mass into
the centers of the potentials.  In any case, most of the gas must
eventually be removed from the proto-dE either by ram-pressure
stripping (Lin \& Faber 1983; Kormendy 1985) and/or by winds from
supernovae and massive stars (Dekel \& Silk 1986; Vader 1986; Yoshii
\& Arimoto 1987).

However, the faded remnant of a dI may appear like the lowest
surface-brightness dEs, often known as dwarf spheroidals (dSph), that
tend to be close companions of large galaxies.  Furthermore, the
bright non-nucleated dEs in Ferguson \& Sandage's (1989) sample of
Virgo and Fornax galaxies show the same spatial distribution as the
dIs, suggesting that the two types may be related.  These lower-mass
systems would also be more easily affected by stripping and winds.  A
connection can also be inferred from the star formation histories of
dEs in the Local Group.  Galaxies like \n147, \n185, Fornax, Sextans,
Leo~I, Sagittarius, and Carina show star formation episodes between 2
and 6 Gyr ago (see compilations by Mateo 1998 and Grebel 1999).  Thus,
they were able to hold onto a significant amount of gas for most of
their lifetimes.

Therefore, determining the evolution of different types of dwarf
galaxies is significant to our understanding of the dwarfs themselves,
group and cluster formation, and the morphology/density relation for
dwarf galaxies.  It would be especially important to study galaxies
that currently have properties of both dEs and dIs.  Examples of such
``transition objects'' in the Local Group are \lgs, Phoenix, Antlia,
DDO~210, and Pegasus (Mateo 1998).  In this paper we report on the
investigation of the star formation history of \lgs using observations
from the \hubble.  The high resolution and resulting photometric depth
of \hst allow the detection of stars below the horizontal branch for
the first time.

Early visual inspection of photographic images of \lgs showed that
something was unusual with its stellar population.  In all previous
cases, gas-rich dwarf galaxies resolved better on blue plates than
visual plates, but the opposite was the case for LGS~3 (Christian \&
Tully 1983).  The deep ground-based photometry of Lee (1995),
Aparicio, Gallart, \& Bertelli (1997, hereafter AGB97), and Mould
(1997) show that the CMD is dominated by a red giant branch (RGB)
extending up to $I\approx20.5$ and $(V\!-\!I)\approx1.3$.  Brighter
and redder than this are asymptotic giant branch (AGB) stars,
including carbon stars (Cook \& Olszewski 1989).  The color and width
of the RGB led Lee (1995) to estimate that [Fe/H]$=-2.1\pm0.2$,
similar to the most metal-poor Galactic globular clusters and dwarf
spheroidal satellites.  The relatively wide giant branch compared to
globular clusters may also be indicative of a spread in age or
metallicity, but photometric uncertainties made this impossible to
determine.  However, unlike these old systems, \lgs does contain
several blue stars that may be young, main-sequence stars or evolved
blue supergiants.  A lack of detectable \hii regions means that there
is no current massive star formation (Hodge \& Miller 1995).

Its stellar population, metallicity, and surface-brightness profile
would have classified \lgs as a dSph.  However, Thuan \& Martin
(1979) detected \ion{H}{1} emission associated with \lgs at a
heliocentric velocity of $-280$~\kms; classifying it as a dI.  Lo,
Sargent, \& Young (1993) confirmed the \hi detection with the
VLA and determined that the total mass of \hi is
$2\times10^{5}~{\rm M}_{\sun}$ and that the dynamical mass is
$(1.8\pm1.0)\times10^{7}~{\rm M}_{\sun}$ for a distance of 760~kpc.
This amount of gas is much lower than normal for a dI galaxy and the
resulting mass to blue light ratio, $26\pm16~{\rm
M}_{\sun}/L_{B,\sun}$, is much higher than is typical for dIs.  The
best distance estimates put it between 770~kpc and 960~kpc
using the $I$ magnitude at the tip of the RGB (Lee 1995; AGB97; Mould
1997).  Lee's surface photometry gives $M_V=-10.35$, making it similar
in luminosity to the Andromeda~III and Sextans dSphs.  Its
distance and location make \lgs a definite member of the Local
Group and it is probably a satellite of either M31 or M33 (Lee 1995).

Cook \etal (1999) have now measured the radial velocities of four
giants in \lgs.  The systemic velocity of $-282\pm4$~\kms is in very good
agreement with the \hi velocity and confirms that the \hi is
associated with the galaxy.  The stellar velocity dispersion of
$7.9_{-2.9}^{+5.3}$~\kms also agrees with the velocity dispersion of the
\hi, and it gives an ``asymptotic'' $M/L$ (corrected for future
fading) $>11$, and possibly as high as 95, similar to other dSphs.

The high resolution of the \hubble now allows us to study the stellar
populations of all Local Group dwarf galaxies to the depth that the
Milky Way satellites could previously be studied from the ground.
Detections of the horizontal branch and the main sequence give a much
better picture of the star formation history of a galaxy than just the
giant branch.  Recent advances in stellar evolutionary models and the
procedures to fit them to color-magnitude diagrams (Tolstoy \& Saha
1996; Aparicio \etal 1996; Dolphin 1997; Dohm-Palmer \etal 1997;
Holtzman \etal 1997; Hernandez, Valls-Gabaud, \& Gilmore 1999) now
allow us to make much more quantitative statements about the star
formation and chemical enrichment histories of resolved galaxies.  We
have used the \hubble to image \lgs in $V$ and $I$ in order to see how
the star formation history of this enigmatic object compares with the
Local Group dwarf spheroidal and dwarf irregular galaxies.  The
observations are described in Section~\ref{sec:obs} and the results
are presented in Section~\ref{sec:results}.  The modeling procedure is
described in Section~\ref{sec:sfhmodel} and some implications are
discussed in Section~\ref{sec:disc}.

\section{Observations and Data Reduction\label{sec:obs}}

\subsection{Observations and Photometry}

Wide Field and Planetary Camera 2 (\wfpc) observations of \lgs were
acquired on 1999 January 9 as part of \hst program GO--6695.  A total
of 18400 seconds of integration (8 orbits) were divided evenly between
the F555W and F814W filters.  Two exposures were taken during each
orbit to aid in the removal of cosmic rays.  The requirements of
avoiding bright stars that could cause either badly saturated pixels
or scattered light, maximizing the scheduling windows, and imaging a
large area of the galaxy resulted in the placement of the galaxy
center on the WF2 detector rather than the higher-resolution PC.
Figure~\ref{fig:voverlay} shows the \wfpc footprint on a ground-based
$V$-image from Lee (1995).  In an effort to gain back some resolution we
employed a 4 position sub-pixel dither observing scheme.  The measured
offsets of the dithers from the F555W PC images were
(0\farcs000,0\farcs000), (0\farcs498,0\farcs253),
(0\farcs761,0\farcs754), and (0\farcs255,0\farcs503) with
uncertainties of $\pm0.002''$. Dithering gave the added advantage of
improving the removal of cosmic rays, hot pixels, and other detector
defects.

To insure the best calibration, the raw images were reprocessed using
the standard pipeline tasks in STSDAS and the best available
calibration files.  The newly calibrated images were then reduced using
three different methods and software to provide internal
checks.  Most of our final results come from using the \hstphot package
(Dolphin 2000a).  

The first step after basic calibration is to combine pairs of images
taken during the same orbit in order to remove cosmic rays.  Two of
the reduction methods used images combined using IDL routines provided
by Dr. Richard White of STScI.  The third method used the \hstphot
utility {\it crclean}.  It incorporates an algorithm similar to that
used in the IRAF task CRREJ; however, it is less prone to rejecting
pixel from real stars in imperfectly registered images.

Photometry was done using aperture photometry on drizzled images and
PSF-fitting on the un-dithered images.  Drizzling combines the images
taken at the four sub-pixel dither positions into an image with higher
effective resolution (Gonzaga \etal 1998).  The drizzling was done
using standard, cookbook procedures with a pixfrac of 0.6.  We were
never able to produce acceptable results with PSF-fitting on the
drizzled images.  However, aperture photometry using methods similar
to those in Miller \etal (1997) worked well in uncrowded regions.
PSF-fitting is required in the crowded regions near the galaxy center.
Therefore, both ALLFRAME (Stetson 1994) and \hstphot were used to do
simultaneous PSF-fitting on the four dither positions.

The \hstphot package is described in detail in Dolphin (2000a), but a
brief description here is warranted since it is a new package and it
is the source of our final photometry.  It is specifically designed to
handle under-sampled WFPC2 images by using a library of TinyTim (Krist
1995) PSFs to account for variations in the PSF due to location on the
chip and the centering within a pixel.  After PSF-fitting corrections
are also made for geometric distortion, CTE effect (Dolphin 2000b),
and the 34$^{th}$ row effect. The four un-drizzled images in each
filter were analyzed simultaneously using the {\it multiphot} routine
to improve the signal-to-noise.  A sharpness parameter and a
measurement of the quality of the fit, $\chi$, are used for
classification and to reject remaining cosmic rays and extended
objects.  Aperture corrections to a 0\farcs5 radius are calculated
from all relatively isolated stars.  The WF chips have many stars and
the aperture correction is well determined.  Unfortunately, there are
not enough appropriate stars on the PC to determine accurate aperture
corrections for that chip.

Artificial star tests were run in order to create a library of
completeness results sampling the range of colors and magnitudes in
this data.  For each 0.5 mag in magnitude by 0.25 mag in color,
approximately 2400 artificial stars were inserted uniformly to give a good
estimate of the completeness fraction and distribution of recovered
photometry.  Figure~\ref{fig:completeness} shows the results from the
completeness tests. The 50\% completeness limits are $V_{50} \approx
27.6$ and $I_{50} \approx 26.6$.

\subsection{Calibration}

Each of the photometry procedures used slightly different zeropoints,
but all of the calibrations used the color terms of Holtzman \etal
(1995).  The aperture photometry used zeropoints determined from
multi-aperture photometry of bright galaxies (Whitmore \etal 1997),
which are very close to the Holtzman \etal (1995) zeropoints. The
ALLFRAME photometry was calibrated using the Hill \etal (1998)
``long-exposure'' zeropoints.  The \hstphot photometry used zeropoints
from Dolphin (2000b) which provides corrections to the Holtzman \etal
(1995) values.  We find that the three photometric methods give very
good agreement in the colors, but that there are systematic offsets in
the magnitudes (Table~\ref{tab:photcomp}).

First, \hstphot magnitudes on the PC are about 0.1 magnitude brighter
than those of stars in the WF chips in similar parts of the CMD. This
is probably due to the larger uncertainties in the PC aperture
correction.  The PC contains only 11\% of the total number of stars
detected and the photometry is not as deep as in the WF chips.
Because of these differences and uncertainties, we do not include the
PC data in the calculation of the star formation history below.  Also,
in the following plots that include PC data, we have added 0.1
magnitude to $V$ and $I$ of the PC points in order to improve the
visual appearance.

For objects detected on the WF chips the agreement is excellent
between \hstphot and aperture photometry.  However, \hstphot gives
magnitudes about 0.1 magnitude brighter in both $V$ and $I$ than
ALLFRAME.  This is probably due to differences in zeropoints and
aperture corrections.  The quality of the ALLFRAME and \hstphot CMDs
in terms of the tightness of the RGB and the number of objects found
is quite similar; both techniques do an equally good job of deblending
stars.  We have chosen the \hstphot results as our standard since they
are based on the most recent calibration (done with the same
software), the good agreement with the aperture photometry, and the
ease of using \hstphot for the artificial star tests needed for the
star formation history analysis.  Tables of the photometry are
available upon request from the author.

\subsection{Comparison with Previous Results}

In Table~\ref{tab:photcomp} we also compare our results with the
previous ground-based work of Lee (1995), AGB97, and Mould (1997).
The \hstphot results agree with Lee (1995) to within 0.02 mag.
However, the \hstphot photometry is 0.07 magnitudes brighter in $V$
and 0.02 magnitudes brighter in $I$ than AGB97.  Thus,
the AGB97 results are closer to our ALLFRAME results.
The Mould (1997) photometry is about 0.25 magnitudes fainter in both
$V$ and $I$ than \hstphot.  Note that in all cases the (\vi) colors
agree within about 5\%.  This uncertainty will have no effect on the
metallicity and star formation history, but it will affect the derived
distance and reddening.

\section{Results\label{sec:results}}

\subsection{The Color-Magnitude Diagram}

The complete color-magnitude diagram (CMD) for \lgs is given in
Figure~\ref{fig:cmd}.  The main features are a narrow RGB, a few AGB
stars just above the RGB, a red clump and blue loop stars, a ``blue
plume'' of main sequence stars, and a blue horizontal branch (HB).
Previous observation had identified the blue loop stars but these are
the first observations to show the blue plume and horizontal branch
clearly.  

\subsection{Distance and Extinction}

The distance can be determined from the $I$ magnitude at the tip of
the red giant branch (TRGB), from the positions of the horizontal
branch and red clump, and from matching the entire CMD using the CMD
modeling technique.  The first three methods require an independent
determination of the reddening while the CMD modeling also fits for
the reddening.

The TRGB method (Lee, Freedman, \& Madore 1993, hereafter LFM93;
Salaris \& Cassisi 1997) is an efficient and widely-used means of
determining distances to resolved galaxies with old, metal-poor
populations.  It makes use of the fact that $M_I$ at the helium flash
(TRGB) is fairly insensitive to age and metallicity.  The $I$
magnitude of the TRGB is usually identified by a step in the
luminosity function.  Unfortunately, our sample of stars is too small
to yield an unequivocal result; an edge-detecting Sobel filter gives a
first peak at $I=20.1\pm0.2$ (Figure~\ref{fig:trgblf}).  An average of
the brightest likely RGB stars gives $I_{TRGB} = 20.12\pm0.26$.  There
is an additional systematic error caused by Poisson statistics.  If
the upper RGB has N stars per magnitude, the brightest star will, on
average, be 1/N mags below the true tip.  In this case, the TRGB is
too faint by $0.07\pm0.06$ mag.  Thus, the TRGB is about 0.5 mag
brighter than used by Lee (1995) and between 0.4 and 1.0 mag brighter
than the two options considered by AGB97.  The color of the TRGB is
$(V-I)=1.50\pm0.05$.

The foreground extinction towards \lgs has been measured to be
$A_B=0.1$ (Burstein \& Heiles 1982), $A_B=0.176\pm0.028$ (Schlegel
\etal 1998), and $A_B=0.14\pm0.06$ (CMD fit, see below). Taking the
mean of the latter two measurements and using $A_B = 4.1E(B-V)$ gives
a mean reddening of $E(B-V)=0.04\pm0.01$.  We assume $R_V=3.1$ and the
reddening law of Mathis (1990) to get $A_B=0.16$, $A_V=0.12$ and
$A_I=0.07$. For comparison, Lee (1995) used $A_V=0.08$, while AGB97
used $A_V=0.12$.  Therefore, the unreddened position of the TRGB is
$I_0=20.12-0.07-0.07 = 19.98\pm0.27$, $(V-I)_0=1.45\pm0.05$.

Continuing with the method of LFM93, the bolometric magnitude of the
TRGB is determined from $M_{bol} = -0.19[{\rm Fe/H}]-3.81$ (Da~Costa \&
Armandroff 1990, hereafter DA90).  The metallicity is calculated from
relations between [Fe/H] and the (\vi) color of the RGB at $M_I=-3.0$
(DA90) and $M_I=-3.5$ (LFM93).  We find $(V-I)_{-3.0,0} = 1.23 \pm
0.05$ and $(V-I)_{-3.5,0}=1.35 \pm 0.03$; giving a mean metallicity of
$[{\rm Fe/H}] = -1.64\pm0.03$ and $M_{bol}=-3.50$.  The absolute
magnitude at the TRGB is $M_I= M_{bol} - BC_I$, where $BC_I = 0.881 -
0.243(V-I)_{TRGB}$ (DA90).  For $BC_I=0.53$, the TRGB has $M_I=-4.03
\pm 0.15$.  The resulting distance modulus is $(m-M)_0 = 24.08 \pm
0.30$.

The horizontal branch is fairly flat and narrow in the $(V,V-I)$
CMD so it can also be used as a standard candle.  Twenty-one stars on
the HB have $\langle V\rangle = 24.71 \pm 0.01$ with an additional
0.05 mag of systematic error due to calibration uncertainties. $M_{V}$
for the HB ranges between $0.5\pm0.2$ and $0.8\pm0.2$.  Dolphin \etal
(2001a) find $M_V=0.61\pm0.08$ for RR Lyraes in IC~1613 whose RGB
color suggests that the metallicity of its old stellar population is
within 0.2 dex of that in \lgs.  Therefore, we adopt the same $M_V$
for the horizontal branch in \lgs. With $A_V=0.12$, this gives a true
distance modulus of $(m-M)_0 = \langle V\rangle - M_{V}(HB) - A_V =
23.98 \pm 0.10$.

The position of the red clump (RC) has also been used as a distance
indicator (Cole 1998; Udalski 2000; Girardi \& Salaris 2001).  One
complication is that the red clump is blended with the red end of the
HB so it is difficult to disentangle the two.  Fitting a quadratic
plus a Gaussian to the luminosity function in the range $0.75 < V-I <
0.95$ and $22<I<24.5$ gives $I(RC) = 23.74 \pm 0.03$.  The
semi-empirical calibration of Dolphin \etal (2001a) and Girardi \&
Salaris (2001) gives $M(RC)_I = -0.29 \pm 0.10$. Therefore, the distance
modulus is $(m-M)_0 = \langle I\rangle - M_{I}(RC) - A_I = 23.96 \pm
0.12$.

Finally, the least-squares solution for the star formation history
(SFH) determines the distance and reddening by fitting all parts of
the CMD simultaneously.  This result is not completely independent of
the three previous methods since the same regions of the CMD are used.
Most of the distance information will come from the positions of the
HB and RC and the distance between the HB and the TRGB. Our best
solution gives $(m-M)_0=23.94\pm0.07$ and $A_V=0.11\pm0.05$,
consistent with the above estimates.  Using our standard extinction of
$A_V=0.12$ gives $(m-M)_0=23.93\pm0.07$.  All the distance estimates
are summarized in Table~\ref{tab:distances}. The result from the SFH
method should be roughly equivalent to averaging results from the
first three methods. The weighted average of the TRGB, HB, and RC
methods is $(m-M)_0=23.98\pm0.07$.  The average of this and the result
from the SFH method then gives $(m-M)_0=23.96\pm0.07$, corresponding
to a distance of $D=620\pm20$~kpc.  We adopt a larger uncertainty than
given by the standard deviation since the errors are correlated.  An
additional source of uncertainty is the 0.1~mag of systematic error in
the zeropoint calibration.

\subsection{Procedure for Modeling the Star Formation History\label{sec:sfhmodel}}

A synthetic CMD-matching technique was used to study the star
formation history (Dolphin 1997; Dolphin 2000c).  The underlying
stellar evolution models are from Girardi \etal (2000) with linear
interpolation in both age and metallicity.  The isochrone spacing is
determinied in order to adequately sample the CMD.  Errors in the
interpolation are between 0.01 dex and 0.05 dex in $\log(t)$, smaller
than the age resolution used here (see Dolphin \etal 2001b).  A
Salpeter initial mass function is assumed.  Dolphin (2000c) contains a
description of the method but we have improved the determination of
the chemical enrichment $Z(t)$. Dolphin (2000c) used multiple fits of
SFR($t$) for each ($(m-M)_0$, $A_V$) pair, each with different $Z(t)$
and assumed $dZ/dt$, to determine the best $Z(t)$ and uncertainty.
Here we make only one fit for each ($(m-M)_0$,$A_V$) combination and
solve for SFR($t$,$Z$).  The step size is 0.05~mag in both distance
and extinction.  This allows solutions for both $Z(t)$ and $dZ(t)$,
the metallicity spread, which the previous method did not allow.
Linear combinations of the CMDs with different ages and metallicities
are then combined, while weighting by SFR(t), to give the final
synthetic CMD.  The youngest age in the fits is 100~Myr since
including younger ages led to a few stars in the synthetic CMD that
were much brighter than any observed stars.  Plotting isochrones on
the observed CMD also supports this decision; there does not seem to
have been any star formation in the last 100~Myr.

The observations are compared with the models in a least-squared sense
using two sets of Hess diagrams (\vi,$V$,Number).  One is
coarsely-binned (0.3 mag magnitude, 0.1 mag color) for sensitivity to
large structures and one is finely-binned (0.15 mag magnitude, 0.1 mag
color) for sensitivity to smaller structures.  Errors and
incompleteness in the artificial CMD are derived from the artificial
star tests.  Rather than assigning errors from a given distribution to
a discreet set of artificial stars, model CMDs are created which
take into account all possible combinations of mass, age, metallicity,
binarity, and photometry. Each cell in the model CMD is weighted by
the completeness from the artificial star tests and then redistributed
according to the distribution of the stars recovered from that
cell. See Dolphin (2001) for a thorough description of the method. The
models are iterated while simultaneously varying $(m-M)_0$, $A_V$, and
SFR($t$,$Z$) to find the combination that minimizes the $\chi^2$
difference.  Because of the large uncertainties in the PC aperture
corrections, we only use data from the WF chips in the modeling.

\subsection{Star Formation History}

\subsubsection{Global Star Formation History}

The star formation rate and metallicity as a function of time for the
global solution to the star formation history are given in
Table~\ref{tab:sfhglobal} and shown in Figs.\ref{fig:sfr}a and
\ref{fig:feh}, where $[{\rm Fe/H}] = \log(Z/0.02)$.  As a whole, \lgs
formed most of its stars very early and has had a slowly decreasing
SFR for the last 10~Gyr.  The procedure for determining the star
formation history from matching the CMD does not make any assumptions
about the chemical evolution and so does not force an increase of
metallicity with time.  Thus, it is possible that we can get the
unphysical situation that the metallicity can decrease with time.
However, the overall enrichment history is a slow and reasonable
increase with time: stars older than 8~Gyr have $[{\rm Fe/H}] =
-1.5\pm0.3$, stars with ages between 1 and 8~Gyr have $[{\rm Fe/H}] =
-1.3\pm0.1$, and stars younger than 1~Gyr have $[{\rm Fe/H}] =
-0.8\pm0.2$, where the averages are weighted by $(SFR{\times}dt)^{-1}$
and the uncertainties represent the metallicity spreads rather than
real standard deviations.

Figure~\ref{fig:hess} shows the high resolution Hess diagram of the
observations and the best-fit global model and an example artificial
CMD from the best-fit model.  The main differences between the
observations and the model are: 1) the synthetic RGB is slightly too
blue (although the red edge of the synthetic RGB falls in right
place); 2) the synthethic RGB is too wide; 3) the RC is too sharp; 4)
the HB is too weak; and 5) the MS is too blue and too strong. Some
systematic effects from these mismatches are that the metallicities
may be slightly off, the SFR at large ages may be underestimated, and
the recent SFR may be overestimated.  However, overall the match is
reasonable with the maximum difference in a single grid cell being
about 3$\sigma$.

\subsubsection{Radial Star Formation Histories}

While the global solution indicates that \lgs is dominated by an old
stellar population, a visual inspection of the photometry suggested
that there are spatial differences in the star formation history.
Therefore, we have also run SFH solutions for three radial zones and
the results are shown in Table~\ref{tab:sfhr} and
Figure~\ref{fig:sfr}b.  In these fits, SFR(t) was solved for while
keeping $Z(t)$, $dZ(t)$, $(m-M)_0$, and $A_V$ fixed at the values from
the global solution.  In Figure~\ref{fig:sfr}b the star formation
rates have been normalized to the rate in the 5--15~Gyr bin in order
to show the relative rates better. The outer zone of the galaxy
($r>76''$) has had a steadily decreasing SFR and no star formation in
the last 200~Myr. The intermediate zone ($21''<r<76''$) has had a
fairly constant SFR between 15~Gyr and 2~Gyr and then a decrease in
the SFR by a factor of 2. The inner $21''$ has had a fairly constant
SFR throughout the lifetime of \lgs that is similar to or slightly
higher than in the other zones.  There has not been significant star
formation anywhere in \lgs in the last 100~Myr.

Color magnitude diagrams for the three zones are shown in
Figure~\ref{fig:cmdrad}.  Isochrones from Girardi \etal (2000) are
over-plotted after being adjusted to a distance modulus of
$(m-M)_0=23.96$ and a reddening of $E(B-V)=0.04$. In the outer zone
(Figure~\ref{fig:cmdrad}a), an extended blue HB from the oldest
population is clearly visible. While most of the stars are older than
2.5~Gyr, a few stars above the HB indicate that there has been some
star formation until about 500~Myr ago.  The intermediate zone
(Figure~\ref{fig:cmdrad}b) has the same old population but a much more
pronounced blue plume from star formation within the last 2~Gyr and a
few blue loop stars that are only 200-300~Myr old.  Finally, the
central $21''$ of the galaxy (Figure~\ref{fig:cmdrad}c) has a higher
ratio of blue to red stars than the other zones, suggesting that the
bulk of the recent star formation has been in the center.  The
magnitudes of the brightest blue loop stars in the inner two zones
suggest that the lower limit on the ages of the most recent stars is
about 100~Myr. Overall, the picture of the SFH from the simple
isochrone fits agrees with the detailed modeling and provides
confidence that the model solution is correct.  The model has the
advantage of being much more quantitative.

\subsubsection{Horizontal Branch Morphology}

In order to look for a second parameter effect in the horizontal
branch, we have measured two indices of horizontal branch type.  The
first is $(N_B-N_R)/(N_B+N_V+N_R)$, where $N_B$, $N_V$, $N_R$ are the
numbers of blue, variable and red HB stars, respectively (Lee,
Demarque, \& Zinn 1994, hereafter LDZ94).  To avoid contamination from
the blue plume, we do this analysis only on stars with projected radii
greater than 76$''$.  There has not been a search for variable stars
in \lgs, so we select the variable candidates as those with $0.35 <
(V-I)_0 < 0.55$ and $23.75 < I_0 < 24.5$.  Only one star is found in
this region of the CMD. The red limit of the HB is taken to be
$(V-I)_0 =0.8$.  With these constraints, the HB type index is
$-0.2\pm0.2$.  This is a lower limit since the number of variable
stars is probably underestimated and there may be some contamination
at the red end of the HB from intermediate-age stars.  We take the
metallicity of the old population to be $[{\rm Fe/H}]=-1.64$ from the
TRGB analysis.  This metallicity is on the Zinn \& West (1984) scale,
allowing us to compare HB type and metallicity with LDZ94.  The
comparison with Figure~7 of LDZ94 shows that \lgs falls on the
sequence of [Fe/H] vs.\ HB type of Galactic globular clusters with
galactocentric distances of between 8 and 40~kpc.  \lgs most resembles
\n6584 or \n7006.

The second HB morphology index is $i=N_B/(N_B+N_R)$ (Da~Costa \etal
1996) where $N_B$ and $N_R$ are the same as defined above.  For \lgs,
the lower limit for $i$ is $0.38\pm0.15$.  The HB of \lgs is bluer
than the HBs of the M31 satellites And~I ($i=0.13\pm0.01$; Da~Costa
\etal 1996) and And~II ($i=0.17\pm0.02$; Da~Costa \etal 2000).
However, it is still redder that the HBs of globular clusters with
similar metallicities.  Da~Costa \etal (2000) calculate that $i=1.0$
for NGC~6752 ($[{\rm Fe/H}] = -1.54$) and M13 ($[{\rm Fe/H}] = -1.65$)
which have metallicities nearly the same as the oldest population in
\lgs.  However, clusters of this metallicity that show a second
parameter effect have much redder HBs: NGC~362 ($[{\rm Fe/H}] =
-1.28$) has $i = 0.04 \pm 0.02$, while NGC~7006 ($[{\rm Fe/H}] =
-1.59$) has $i = 0.23 \pm 0.06$.  Therefore, there is evidence for a
moderate second parameter effect in \lgs.  Again, we stress the caveat
that HB morphologies in galaxies with more than simple stellar
populations such as And~I, And~II, and \lgs are uncertain because of
possible contamination of the red side of the HB from younger
populations.

\subsubsection{Clusters of Young Stars}

It is also informative to look at the spatial distributions of stars
in various parts of the complete CMD.  For example,
Figure~\ref{fig:hbspat} shows the very even distribution of the oldest
stars on the HB.  This is the most uniform of any population.
Alternatively, the main sequence stars are the most unevenly
distributed (Figure~\ref{fig:msspat}). The young stars are concentrated
towards the center of \lgs and are clearly grouped into clusters or
associations. We have identified possible clusters using the Path
Linkage Criterion (Battinelli 1991) on stars in WF2 with $I<26$ and
$(\vi)<0.3$. Thirteen groups with more than 5 stars are found with a
characteristic scale of $d_s=4.1$~pc.  The clusters are shown in
Figure~\ref{fig:wf2groups}.  The properties of the clusters are given
in Table~\ref{tab:wf2groups}.  The crosses in each region of
Figure~\ref{fig:wf2groups} are the centroids of the groups and the
sigmas of the centroids in x and y.  These sigmas give characteristic
sizes of 2--5~pc, consistent with $d_s$. Table~\ref{tab:wf2groups}
also gives the distances between the most widely separated stars in
each group. The largest have diameters of about 25~pc.
Figure~\ref{fig:cmdclust} shows the CMD of the stars in the largest of
these clusters, Cluster 9, and a comparably sized background region.
The cluster has a blue plume, a weak RGB, and a red HB.  Comparison
with isochrones suggests that the cluster is on the order of 800~Myr
old.

\section{Discussion\label{sec:disc}}

\subsection{Distance}

The new distance modulus of $(m-M)_0=23.96\pm0.07$ (620~kpc) is
considerably smaller than the previous values of $24.5\pm0.2$
(810~kpc; Lee 1995), $24.4\pm0.2$ (770~kpc; AGB97), and $24.6\pm0.4$
(830~kpc; Mould 1997).  While our TRGB distance is rather uncertain,
use of the HB and the RC should make our distance the most reliable to
date.  While the new distance requires adjustments to most of the
physically derived parameters of \lgs, it does not radically alter the
character or location of the galaxy.  The observed structural and
kinematic properties and updated derived physical parameters are given
in Table~\ref{tab:properties}.

\lgs is located 20\degr\ from M31 and 11\degr\ from M33, and the three
galaxies have systemic velocities with respect to the center of the
Local Group of $-59$~\kms, $-26$~\kms, and 48~\kms, respectively
(Schmidt \& Boller 1992).  This suggests that \lgs could be a
satellite of either M31 or M33 (Lee 1995).  Using distances to M31 and
M33 of 770~kpc and 870~kpc (Mould \& Kristian 1986; LFM93), we find
that \lgs is 280~kpc from each.  The main difference compared to
previous results is that \lgs is $\sim100$~kpc further away from M33.
This, the larger mass of M31, and the better agreement in the
velocities between \lgs and M31, suggest that if \lgs is a satellite,
it is probably orbiting M31.

Of course, the consequences of \lgs being closer are that it is
smaller and fainter.  The absolute magnitude within an aperture radius
of 106$''$ is now $M_V=-9.82\pm0.09$, the core radius is
$R_c=148\pm9$~pc, and the most likely dynamical mass is
$M=(2.6^{+4.7}_{-1.6})\times10^7$~M$_{\sun}$.  The quantities
described as ``most likely'' are related to the most likely value of
the stellar velocity dispersion, $\sigma = 7.9_{-2.9}^{+5.3}$~\kms,
from Cook \etal (1999).  However, since $M/L=\rho_0/I_0 \propto
\sigma^2/I_{0}R_{1/2}$, where $\sigma$ is the velocity dispersion and
$R_{1/2}$ is the radius where the surface brightness falls to one-half
of the central value (Richstone \& Tremaine 1986), the central density
and mass-to-light ratio increase. The integrated $M/L$ for the most
likely value of the velocity dispersion is $(M/L)_V=36^{+64}_{-22}$,
while the asymptotic $M/L$, assuming that \lgs will fade by a factor
of 3.9, is $(M/L)_{V,a}=140^{+252}_{-84}$, compared to
$(M/L)_{V,a}=95^{+175}_{-56}$ using a distance of 810~kpc (Cook \etal
1999).  This gives \lgs the highest asymptotic $M/L$ of any Local
Group galaxy (see Figure~4 of Cook \etal 1999).  Either the dynamics
are completely dominated by dark matter, the dynamics are
non-Newtonian (cf. Cook \etal 1999), or it is not virialized.

\subsection{Comparison of the Stellar Populations and the HI distribution}

\lgs is also rather unique in the Local Group in that it has a very
smooth \hi distribution that is centered on the optical galaxy
(Figure~\ref{fig:lgs3vhi}).  Figure~\ref{fig:wfpc2hi} shows the \hi
distribution superposed on the WFPC2 mosaic. Most other star forming
dwarf irregular galaxies (e.g. SagDIG, Young \& Lo 1997, hereafter
YL97; IC~10, Wilcots \& Miller 1998) have \hi holes that may be
created by stellar winds or supernovae.  Leo~A also has a fairly
symmetric \hi distribution, but, like many star forming dwarfs, it has
both narrow ($\sigma\sim5$~\kms) and broad ($\sigma\sim10$~\kms)
components (Young \& Lo 1996).  The narrow, cold component is
associated with areas of current star formation.  In \lgs, the line
profiles are fit by single Gaussians with widths between 5 and 10~\kms
with a median dispersion of 8.5~\kms (YL97).  Some of the dwarf
spheroidals such as Phoenix and Sculptor may also have associated \hi
(Carignan, Demers, \& C\^ot\'e 1991; Carignan \etal 1998; Blitz \&
Robishaw 2000) but those clouds are found at large projected distances
from the galaxies.  

The properties of the gas in \lgs are consistent with a lack of star
formation in the last $\sim100$~Myr.  Gas moving at 5~\kms can travel
about 500~pc, or over 3 core radii and a distance larger than the
diameter of most \hi holes, in 100~Myr.  So, any \hi holes that might
have been formed during the last period of significant star formation
could have been filled in.  The apparent lack of a cold phase to the
ISM in \lgs suggests that the pressures are too low to contain a large
amount of high density gas for forming star (YL97).  Likewise, most of
the gas in \lgs is below the gravitational instability threshold.
Assuming a thin, isothermal disk, a flat rotation curve, and a
velocity dispersion of 6~\kms, the disk will be unstable if the gas
density exceeds
\begin{equation}
\Sigma_c(M_{\sun} \ {\rm pc}^{-2}) = 0.59\alpha V ({\rm km \
s}^{-1})/R({\rm kpc})
\end{equation}
where $V$ is the rotation velocity and $\alpha$ is a dimensionless
geometrical parameter with a value of 0.7 for spiral galaxies
(Kennicutt 1989).  At a radius of 76$''$ (228~pc), the critical
density for $V=5$~\kms is $\Sigma_c=9$~$M_{\sun}$~pc$^{-2}$.  The \hi
column densities at this radius are $(5-10)\times10^{19}$~cm$^{-2}$,
or $\Sigma=0.4-0.8$~$M_{\sun}$~pc$^{-2}$, a factor of 10 below the
critical density.  It must be noted that the resolution of the \hi
data is 35$''$, so there could be small pockets of dense gas where
star formation could proceed slowly.

One must also be concerned about whether the star formation that we
measure is sufficient to expel all the gas from the galaxy.  Young \&
Lo (1997) calculate that a $10^4$~$M_{\sun}$ burst that produces 100
supernovae would provide enough energy to blow all the gas out of the
galaxy.  Our star formation history solution gives only
$\sim600~M_{\sun}$ of stars forming in the last period of
measurable star formation between 100 and 200~Myr ago.
Even if all these stars formed simultaneously, this is much less than
the amount needed to drive out the gas.  Also, there is no other
recent burst sufficiently strong to expel the current \hi mass.

Since stellar winds and supernovae have had insufficient energy input
to expel gas from \lgs, much of the enriched gas from stellar
evolution has remained in the galaxy.  In fact, a large fraction of
the current gas may have been processed through stars.  The
Starburst99 models for continuous star formation and $Z=0.001$ give an
asymptotic mass-loss rate of $0.25\times SFR$~\msun yr$^{-1}$
(Leitherer \etal 1999).  The time averaged star formation rate for
\lgs is $\sim5\times10^{-5}$ \msun yr$^{-1}$.  Therefore, over 15~Gyr
the stars of \lgs have expelled about $2\times10^5$ \msun of gas, or a
little over half of the current total gas mass
(Table~\ref{tab:properties}).  This is also consistent with the
chemical evolution calculations of AGB97 which showed that for their
lower distance of 770~kpc the chemical history was best fit by a zero
or low outflow model.  Thus, \lgs is close to being the ideal
closed-box system.

\subsection{Star Formation History}

The star formation history of \lgs is dominated by a very early burst
followed by a low level of continuous star formation in the center.
The SFH of the outer region is similar to the other Local Group dwarf
elliptical and dwarf spheroidal galaxies such as Sculptor, And~I,
And~III, Sextans, Tucana, Draco, Ursa Minor, Phoenix, and NGC~205
(Mateo 1998).  However, the continued and/or rejuvenated star
formation in the center is more typical of some of the low-mass dwarf
irregular galaxies and larger dEs like Phoenix, Fornax, WLM, Antlia,
and NGC~205.  This stresses the similarities that \lgs has always had
with both dE/dSph and dI galaxies.

The Phoenix dwarf is on the lists of both dE/dSph and dI galaxies with
star formation histories similar to \lgs.  Like \lgs, Phoenix has been
classified as a dSph/dI galaxy (Mateo 1998).  Early ground-based CMDs
showed it to be dominated by old stars with a small population of
young, blue stars (Canterna \& Flower 1977; Ortolani \& Gratton 1988;
van de Rydt, Demers, \& Kunkel 1991) and there have been various
reports of associated \hi (Carignan \etal 1991; Young \& Lo 1997;
St-Germain \etal 1999).  Recent deep color-magnitude diagrams of
Phoenix (Holtzman \etal 2000; Held \etal 1999; Mart{\'\i}nez-Delgado
\etal 1999) combined with the results presented here, only enforce the
similarities.  The WFPC2 CMD of Phoenix (Holtzman \etal 2000) is
strikingly similar to Figure~\ref{fig:cmd}.  The main difference is
that Phoenix has a more pronounced blue plume of young stars.  Using
the calibration of Da~Costa \& Armandroff (1990) gives metallicities
for the old, giant branch stars of $-1.9<[{\rm Fe/H}]<-1.45$, similar
to that found for \lgs using the same technique and calibration.  The
overall enrichment histories of the galaxies are similar, with most of
the younger stars having $[{\rm Fe/H}] < -1$.  However, about 5\% of
the stars in Phoenix have $[{\rm Fe/H}] > -0.5$, more metal-rich than
any of the stars in \lgs.  Overall, however, they are more alike than
different.

As might be expected from these similarities, the star formation
histories of \lgs and Phoenix have many common features.  The bulk of
their stars formed more than 8~Gyr ago and, like in \lgs, the young
stars in Phoenix are concentrated in the center (Mart{\'\i}nez-Delgado
\etal 1999; Held \etal 1999).  Holtzman \etal (2000) computed the star
formation history for the central region of Phoenix using a method
similar to that employed here.  Confirming earlier work, they find that
the star formation rate in the center has been fairly uniform.
However, comparison with Figure~\ref{fig:sfr} suggests that Phoenix
has had larger variations of SFR than has \lgs and that Phoenix has a
higher fraction of stars with ages less than 3~Gyr, thus the more
prominent blue plume.  However, both datasets need to be modeled using
the same method before very detailed comparisons can be made.

The concentration of young stars toward the center of dwarf galaxies
is quite common (see Lee 1993; Minniti \& Zijlstra 1996; Mateo 1998).
Currently it is not known whether there is a kinematic halo/disk
structure like in the Galaxy or whether this is a pure population
gradient.  Mart{\'\i}nez-Delgado \etal (1999) mention several reasons
for why this could happen in Phoenix: 1) all outer stars were formed
in a short initial burst; 2) star formation ceased a few Gyr ago in
the outer regions; and 3) star formation has been continuous but very
low in the outer parts.  Our SFH solutions suggest a fourth option, a
more rapidly declining star formation rate with time in the outer
region compared to the center. This is similar to options 2 and 3 in
that star formation in the outer region seems to have stopped
recently, but not a few Gyr ago, and that star formation in the outer
region over the last 2 Gyr has been fairly flat but slowly declining.
This assumes that the stars were formed at radii similar to where we
find them today.  A large early burst of star formation occurred when
the gas supply was large and extended, forming a broad sheet of
metal-poor stars.  As the gas supply is used up or stripped off in the
outer parts, the star formation rate drops and becomes undetectable
200-500 Myr ago.  Star formation becomes concentrated towards the
center.  Over most of the galaxy the gas density is below the critical
threshold, but random motions, stellar winds, or weak interactions (if
it is indeed orbiting M31) could occasionally cause small regions of
gas to exceed the critical density and collapse in the central
regions.

An alternative scenario is that all the star formation has occurred in
the central region where the gas density will be the highest and most
likely to exceed the critical threshold.  Since the velocity dispersion
of \lgs is on the order of any possible rotational velocity, it is
conceivable that simple proper motions could be responsible for older
stars being found at larger radii.  However, simple dynamical
considerations argue against this.  The time for a star moving at
8~\kms to traverse the 300~pc radius of the galaxy is $t_{cross}
= 3.7\times10^7$~years.  Thus, stars from the most recent
episode of star formation 200~Myr ago have had sufficient time to
diffuse throughout the galaxy if their energies permitted it.  The
fact that we still see a population gradient suggests that the orbits
of the young stars are confined to the central region of \lgs.

While \lgs is on the order of 400 crossing times old, this is
insufficient time for star-star interactions to significantly affect
its structure.  The relaxation time, or time when stellar encounters
become important, is given by
\begin{equation}
t_{relax} = \frac{0.1N}{\ln N}t_{cross}
\end{equation}
(Binney \& Tremaine 1987). If \lgs contains $\sim10^6$ stars, then
$t_{relax} \approx 2.7\times10^{11}$~years, or much longer than a
Hubble time.  Also, dispersion-dominated collisions between stars and
gas clouds are unlikely to be important because individual cloud cores
will have low masses and because the relative velocities between stars  
and clouds are likely to be small.  Therefore, the current locations
of the stars reflect the radii where they were formed.

Overall, the star formation histories of low-luminosity dwarf
irregular and dwarf spheroidal galaxies are remarkably similar: the
SFH is dominated by an early burst followed by low levels of star
formation in the center.  The main difference between dSphs like Draco
and galaxies like \lgs or GR8, is the amount of gas that they can
retain for future star formation.  This may be highly dependent on
environment.  \lgs, Phoenix, and many of the dIs are found throughout
the Local Group and may not be satellites of a larger spiral.
However, most of the dwarf spheroidals without large amounts of
associated gas are found within 250~kpc of either M31 or the Galaxy.
Blitz \& Robishaw (2000) argue that ram-pressure stripping from the
hot halos of M31 and the Galaxy can explain the loss of gas.  They
also find a more extended \hi cloud 10.8~kpc from \lgs that is offset
from the detection of Young \& Lo (1997) by 50~\kms and that could be
due to ram-pressure stripping, but the velocity offset has the
opposite sign to what stripping should produce.  The \hi gas
associated with Phoenix may also be offset from the body of the galaxy
due to ram-pressure stripping (Gallart \etal 2001).

\section{Conclusions\label{sec:conc}}

We have determined a new distance, reddening, and star formation
history for \lgs from \hst observations.  Our main results are:

\begin{enumerate}
\item We have resolved the horizontal branch and the young main
sequence for the first time.
\item The HB type index of $-0.2\pm0.2$ puts \lgs on the sequence of
[Fe/H] vs. HB type of Galactic globular clusters with galactocentric
distances between 8 and 40~kpc.  The HB is bluer than those of outer
halo globulars or M31 dwarf spheroidal satellites.
\item The mean metallicity has increased from $[{\rm Fe/H}] =
-1.5\pm0.3$ for stars older than 8~Gyr to $[{\rm Fe/H}] \approx -1$
for the most recent generation. 
\item The new distance measured from the TRGB, the horizontal
branch, and the red clump is $D=620\pm20$~kpc.
\item We have determined the star formation history of \lgs by
quantitatively fitting the color-magnitude diagram with evolutionary
models.  The global star formation history is dominated by a burst at
$13-15$~Gyr and has been fairly constant since then.  However, there
is clear evidence for a radial dependence on the SFH. Star formation 
has been constant in the central 21$''$ while in
the outer regions ($r>76''$) the star formation rate has been
declining with time.  Dynamical arguments suggest that the current
locations of the stars reflect the radii where they formed.
\item There have not been sufficient bursts of star formation to expel
the gas from the galaxy.  Thus, a large fraction of the current \hi
could have been produced by stellar mass loss.
\item The closer distance makes \lgs less massive but more dense so
the $M/L$ ratios are in the range $3.9<(M/L)_{\sun V}<535$ with a most
likely value of $(M/L)_{\sun V}=36^{+64}_{-22}$.
\end{enumerate}

Whether classified as irregular or spheroidal, galaxies like \lgs are
important for studying the evolution of dwarf galaxies because they
are the lowest-mass galaxies that have managed to retain some \hi and
have star formation until very recently.  Also, they appear to be
nearly ideal closed-box systems.  Further work to help clarify how
star formation proceeds in \lgs would include higher resolution \hi
mapping, deep, wide-field imaging to determine the full extent of
\lgs and whether it has extra-tidal stars, and more radial velocity
measurements to look for changes in velocity dispersion with radius.

\acknowledgements

We would like to thank Lisa Young for providing us with \hi maps of
\lgs, Paulo Battinelli for use of the PLC code, and the referee for
useful suggestions.  This research has made use of NASA's Astrophysics
Data System Abstract Service and Extragalactic Database (NED).  BWM
would like to thank the Carnegie Institution of Washington, the Space
Telescope Science Institute, and the University of Leiden for support
during this project. This research was supported by the Gemini
Observatory, which is operated by the Association of Universities for
Research in Astronomy, Inc., under a cooperative agreement with the
NSF on behalf of the Gemini partnership: the National Science
Foundation (United States), the Particle Physics and Astronomy
Research Council (United Kingdom), the National Research Council
(Canada), CONICYT (Chile), the Australian Research Council
(Australia), CNPq (Brazil) and CONICET (Argentina).  MGL is supported
in part by the MOST/KISTEP International Collaboration research grant
(99-01-009).  Finally, we are grateful for support from NASA through
grants GO-6695, GO-7496, and GO-2227.06-A from the Space Telescope
Science Institute, which is operated by AURA, Inc., under NASA
contract NAS5-26555.

\clearpage

\begin{figure}
\caption{An overlay of the \wfpc footprint on a
ground-based $V$-band CCD image of \lgs from Lee (1995).  The center of the
galaxy was placed on the WF2 detector in order to maximize coverage of
the galaxy while avoiding the nearby bright
stars.\label{fig:voverlay}}
\end{figure}

\begin{figure}
\caption{Results of artificial-star completeness
tests. The 50\% completeness limits are $V\approx27.6$ and $I\approx
26.6$.\label{fig:completeness}}
\end{figure}

\begin{figure}
\caption{Color-magnitude diagram of all stars detected by WFPC2 in
\lgs. The main features are the narrow RGB, a horizontal branch that
extends to the blue, a blue plume of young stars, and blue loop and
red-clump stars from an intermediate-age population.\label{fig:cmd}}
\end{figure}

\begin{figure}
\caption{The luminosity function of stars in the upper
part of the RGB.  The small sample size prevents there being a large
jump in the LF at the TRGB.  As a result, application of a Sobel
filter (dotted line), does not produce a strong peak where the TRGB is
likely to be.  It is most likely that the TRGB has $I\approx20$ or
$I\approx 20.5$.\label{fig:trgblf}}
\end{figure}

\begin{figure}
\caption{The star formation rate as function of
time as derived from the least-squares fits to the CMD. a) The global
solution (Table~\ref{tab:sfhglobal}). b) Solutions for three radial
bins, normalized to the SFR in the 5--15 Gyr bin
(Table~\ref{tab:sfhr}).  The different binning is produced by
averaging the previous bins rather than running new solutions with
different bin sizes.\label{fig:sfr}}
\end{figure}

\begin{figure}
\caption{The global metallicity as a function of
time from the least-squares fits to the CMD.\label{fig:feh}}
\end{figure}

\begin{figure}
\caption{a-d) High resolution (e-h) Hess diagrams
showing the comparison between the observed and modeled CMDs. 
a) observations; b)
best-fit synthetic; c) residuals (light = more observed; dark = more
synthetic); d) residuals in terms of sigma, the darkest shades
correspond to a 3$\sigma$ difference. e) An example synthetic CMD from
the best model. \label{fig:hess}}
\end{figure}

\begin{figure}
\caption{CMDs of stars in different radial bins with isochrones from 
Girardi \etal (2000) over-plotted.  ($Z$,$\log(t)$) for the models are
(0.001,8.3),  (0.001,8.8), (0.001,9.4), (0.001,10.1), and
(0.0004,10.2). The models have been shifted to match the data using
 $(m-M)_0=23.96$ and $E(B-V)=0.04$. a) $r>76''$; b) $21'' < r < 76''$;
and c) $r<21''$.\label{fig:cmdrad}}
\end{figure}

\begin{figure}
\caption{CMD and spatial distribution of the horizontal
branch stars.  These stars are the most uniformly distributed type
in \lgs.\label{fig:hbspat}}
\end{figure}

\begin{figure}
\caption{CMD and spatial distribution of the main sequence stars showing clear
groupings near the center of the galaxy.\label{fig:msspat}}
\end{figure}

\begin{figure}
\caption{Clusters of blue stars found in WF2 using the PLC method
(Battinelli 1991).  Circled stars are the members of the clusters
and the boundaries are drawn by eye to help distinguish them.  Crosses
give the centroids of the clusters and the sigmas in x and y.  
The PLC scale length is $d_s=4.1$~pc, but the
largest linear dimension of a cluster can be up to 27~pc 
(Table~\ref{tab:wf2groups}).\label{fig:wf2groups}}
\end{figure}

\begin{figure}
\caption{a) CMD of stars in Cluster 9, the largest of the young
stellar groupings (Table~\ref{tab:wf2groups}).  Over-plotted is an
isochrone from Girardi \etal (2000) with $Z=0.001$, $\log(t)=8.9$,
$(m-M)_0=23.96$, and $E(B-V)=0.04$.  The match suggests that the age
of the cluster is about 800~Myr. b) A CMD of region of WF2 of
comparable size to that used in (a), but not containing any
clusters. The isochrone is for $Z=0.001$, $t=3.5$~Gyr and the good
match suggests that most of the stars in this region are older than a
few Gyr.\label{fig:cmdclust}}
\end{figure}

\begin{figure}
\caption{Contours of \hi column density from Young \& Lo (1997)
superposed on the ground-based $V$ image from Lee (1995).  Unlike
other dI galaxies with gas, the \hi has
a smooth distribution that follows the stellar distribution. Contour
levels are 2.5, 5.0, 7.5, 10.0, and
$12.5\times10^{19}$~cm$^{-2}$.\label{fig:lgs3vhi}}
\end{figure}

\begin{figure}
\caption{Contours of \hi column density from Young \& Lo (1997)
superposed on the WFPC2 image.  The most recent star formation and the
central clusters are near the region of highest \hi column density. 
Contour levels are 2.5, 5.0, 7.5,
10.0, and $12.5\times10^{19}$~cm$^{-2}$.\label{fig:wfpc2hi}}
\end{figure}

\clearpage
\begin{deluxetable}{lcccl}
\tablewidth{0pt}
\tablecaption{Photometry Comparisons\label{tab:photcomp}}
\tablehead{
\colhead{Detector} & \colhead{$\delta V$} & \colhead{$\delta(V-I)$} &
\colhead{$\delta I$} & \colhead{Comments}
}
\startdata
\sidehead{\hstphot$-$ALLFRAME}
   PC  &  $-0.23\pm0.04$ & $-0.02\pm0.04$ &  $-0.22$ &  $20<V<25$, $20<I<24$\\
   WF  &  $-0.09\pm0.02$ & $0.00\pm0.01$  &  $-0.09\pm0.02$ & \\
\sidehead{\hstphot$-$Aperture}
   PC  &  $-0.14\pm0.03$ & $+0.06\pm0.03$ &  $-0.20\pm0.03$ &  $V<25$\\
   WF  &  $+0.00\pm0.06$  & $+0.03\pm0.05$ &  $-0.03\pm0.06$ & \\
\sidehead{\hstphot$-$Lee(1995)}
   WF2 &  $+0.02\pm0.11$  & $+0.02$        &  $+0.00\pm0.16$ &
$20<V<22.8$, $19<I<22$\\
\sidehead{\hstphot$-$Aparicio \etal (1997)}
   WF  &  $-0.07$        & $-0.05$        &  $-0.02$  & \\ 
\sidehead{\hstphot$-$Mould (1997)}
   WF  &  $-0.24$        & $0.02$         &  $-0.26$  & \\
\enddata
\end{deluxetable}

\begin{deluxetable}{cccc}
\tablewidth{0pt}
\tablecaption{Summary of Distance Estimates\label{tab:distances}}
\tablehead{
\colhead{Method} & \colhead{$E(B-V)$} & \colhead{$(m-M)_0$} & \colhead{Distance}\\
 & & & (kpc)
}
\startdata
TRGB     & 0.04 & $24.08 \pm 0.30$ & $655\pm90$\\
HB       & 0.04 & $23.98 \pm 0.10$ & $625\pm30$\\
RC       & 0.04 & $23.96 \pm 0.12$ & $620\pm30$\\
SFH      & 0.04 & $23.93 \pm 0.07$ & $610\pm20$\\
\enddata
\end{deluxetable}

\begin{deluxetable}{cccc}
\tablewidth{0pt}
\tablecaption{Global SFH\label{tab:sfhglobal}}
\tablehead{
\colhead{Age Range} & \colhead{SFR} & \colhead{mean [Fe/H]} & \colhead{$\sigma$([Fe/H])}\\
(Gyr) & (\e{-5} \msun yr\uu{-1}) & & 
}
\startdata
 0.1 -- 0.2 & $0.6\pm0.2$ & $-1.25\pm0.50$  & $0.26\pm0.31$\\
 0.2 -- 0.4 & $0.9\pm0.2$ & $-0.90\pm0.08$  & $0.42\pm0.03$\\
 0.4 -- 0.6 & $1.8\pm0.3$ & $-0.93\pm0.07$  & $0.35\pm0.08$\\
 0.6 -- 1.0 & $1.6\pm0.4$ & $-0.74\pm0.10$  & $0.24\pm0.09$\\
 1.0 -- 2.0 & $1.3\pm0.6$ & $-1.24\pm0.26$  & $0.50\pm0.13$\\
 2.0 -- 5.0 & $2.3\pm0.8$ & $-1.24\pm0.15$  & $0.45\pm0.09$\\
 5.0 -- 8.0 & $1.7\pm1.2$ & $-1.34\pm0.22$  & $0.18\pm0.15$\\
 8.0 -- 11.0 & $1.2\pm1.0$ & $-1.70\pm0.25$ & $0.21\pm0.16$\\
11.0 -- 13.0 & $1.5\pm0.8$ & $-1.94\pm0.40$ & $0.22\pm0.15$\\
13.0 -- 15.0 & $26.6\pm6.2$ & $-1.32\pm0.10$ & $0.31\pm0.07$\\
\enddata
\tablecomments{$A_V=0.11$, $(m-M)_0=23.94$, ${\rm IMF\ slope} = 1.35$}
\end{deluxetable}

\begin{deluxetable}{cccc}
\tablewidth{0pt}
\tablecaption{Radial SFH\label{tab:sfhr}}
\tablehead{
\onec{Age Range} & \multicolumn{3}{c}{SFR/SFR(10Gyr)}\\
\colhead{(Gyr)} & \colhead{$r<21''$} & \colhead{$21''<r<76''$} &
\colhead{$r>76''$}
}
\startdata
 0.1 --  0.2 &  $1.64 \pm 0.36$ &  $0.62 \pm 0.46$ &  $0.00 \pm 0.00$\\
 0.2 --  0.4 &  $0.91 \pm 0.27$ &  $0.27 \pm 0.08$ &  $0.05 \pm 0.05$\\
 0.4 --  0.6 &  $1.82 \pm 0.73$ &  $0.58 \pm 0.04$ &  $0.14 \pm 0.05$\\
 0.6 --  1.0 &  $0.73 \pm 0.09$ &  $0.23 \pm 0.08$ &  $0.10 \pm 0.05$\\
 1.0 --  2.0 &  $0.73 \pm 0.18$ &  $0.54 \pm 0.12$ &  $0.14 \pm 0.05$\\
 2.0 --  5.0 &  $1.55 \pm 0.36$ &  $1.08 \pm 0.04$ &  $0.67 \pm 0.19$\\
 5.0 -- 15.0 &  $1.00 \pm 0.09$ &  $1.00 \pm 0.08$ &  $1.00 \pm 0.05$\\
\enddata
\end{deluxetable}

\clearpage
\begin{deluxetable}{rrccccr}
\tablewidth{0pt}
\tablecaption{Clusters with $V-I<0.3$, $I<26$, $d_s=4.1$~pc, and $N_{star}>5$\label{tab:wf2groups}}
\tablehead{
\colhead{Id} & \colhead{$N_{star}$} & \colhead{RA} &
\colhead{Dec} & \colhead{$\sigma(x)$} & \colhead{$\sigma(y)$} & 
\colhead{$D_{max}$}\\
 & & (J2000) & (J2000) & [pc] & [pc] & [pc]
}
\startdata
  1 &	21 &  01:03:56.185 &  21:52:58.06 &   4.4 &  3.2 & 19.8\\
  2 &	 5 &  01:03:55.068 &  21:52:41.66 &   2.4 &  3.0 &  9.4\\
  3 &	 8 &  01:03:55.602 &  21:52:54.55 &   3.7 &  4.3 & 14.3\\
  4 &	 9 &  01:03:54.916 &  21:53:01.07 &   3.5 &  1.4 & 11.5\\
  5 &	21 &  01:03:55.752 &  21:53:17.43 &   3.9 &  5.3 & 23.4\\
  6 &	 5 &  01:03:53.704 &  21:52:31.45 &   2.2 &  2.3 &  6.8\\
  7 &	 9 &  01:03:55.325 &  21:53:07.36 &   2.6 &  1.1 &  8.6\\
  8 &	 6 &  01:03:55.347 &  21:52:32.48 &   3.6 &  1.4 &  9.6\\
  9 &	33 &  01:03:54.320 &  21:52:43.80 &   4.3 &  5.4 & 26.9\\
 10 &	15 &  01:03:54.653 &  21:53:19.51 &   2.8 &  4.2 & 14.2\\
 11 &	 6 &  01:03:54.000 &  21:53:07.43 &   3.4 &  0.9 &  8.6\\
 12 &	 7 &  01:03:53.674 &  21:52:50.85 &   2.9 &  2.1 &  8.1\\
 13 &	 5 &  01:03:55.801 &  21:52:27.17 &   2.1 &  2.1 &  6.6\\
\enddata
\end{deluxetable}

\clearpage
\begin{deluxetable}{lcll}
\tablewidth{0pt}
\tablecaption{General Properties of \lgs.\label{tab:properties}}
\tablehead{
\colhead{Parameter} & \colhead{Value} & \colhead{Units} & \colhead{Source}
}
\startdata
\cutinhead{Observed}
Distance modulus     & $23.96\pm0.07$  & mag & 1\\
Distance             & $620\pm20$    & kpc & 1\\
Foreground reddening & $0.04\pm0.01$ & mag & 1,2\\
$V$ integrated ($r<106''$) & $14.26\pm0.04$ & mag & 3\\
Core radius          & $0.82\pm0.05$ & arcmin & 3\\
Ellipticity          & $0.2\pm0.1$   &        & 3\\
$V$ central surface brightness & $24.8\pm0.1$ & mag arcsec$^{-2}$ & 3\\
Central velocity dispersion & $2.6-30.5$ & \kms & 4\\
``Most likely'' velocity dispersion & $7.9^{+5.3}_{-2.9}$ & \kms & 4\\
Optical systemic velocity & $-282\pm4$ & \kms & 4\\
HI systemic velocity & $-286.5\pm0.3$ & \kms & 5\\
Integrated HI flux   & $2.7\pm0.1$ & Jy \kms & 5\\
\cutinhead{Derived}
Absolute magnitude ($M_V$)  & $-9.82\pm0.09$        & mag & \\
Core radius          & $148\pm9$     & pc  & \\
Central surface brightness ($S_{0,V}$) & $4.4\pm0.8$   & $L_{\sun V}$ pc$^{-2}$ & \\
Central mass density ($\rho_0$)   & $0.05-6.9$    & $M_{\sun}$ pc$^{-3}$ & \\
Total mass           & $2.8\times10^6 - 3.9\times10^8$ & $M_{\sun}$ & \\
``Most likely'' mass & $(2.6^{+4.7}_{-1.6})\times10^7$ & $M_{\sun}$ & \\
Central luminosity density & $0.014\pm0.003$ & $L_{\sun V}$ pc$^{-3}$ & \\
Central $M/L$        & $3.6-490$    & $M_{\sun}/L_{\sun V}$ & \\
Total luminosity     & $(7.2\pm0.6)\times10^5$ & $L_{\sun V}$ & \\
Integrated $M/L$     & $3.9-535$ &  $M_{\sun}/L_{\sun V}$ & \\
``Most likely'' $M/L$ & $36^{+64}_{-22}$  & $M_{\sun}/L_{\sun V}$ & \\
Asymptotic $M/L$     & $140^{+252}_{-84}$  & $M_{\sun}/L_{\sun V}$ & \\
Total gas mass (incl. He) & $(3.4\pm0.2)\times10^5$ & $M_{\sun}$ & \\
Gas mass fraction    & $9\times10^{-4} - 0.1$ & & \\
\enddata
\tablerefs{(1) This work; (2) Schlegel \etal 1998; (3) Lee 1995; (4) Cook \etal
1999; (5) Young \& Lo 1997}
\end{deluxetable}

\end{document}